\documentclass[conference]{IEEEtran}  
\IEEEoverridecommandlockouts                              

\usepackage{graphicx} 
\usepackage{subcaption}
\usepackage{cite}
\usepackage{bm}
\usepackage{mathtools}
\mathtoolsset{showonlyrefs}
\usepackage[ruled,vlined]{algorithm2e}

\newcommand{\rev}[1]{{\color{black} #1}}
\usepackage{amsmath, amssymb, amsfonts, enumerate, amsthm}
\usepackage{mathtools}

\providecommand{\argmax}{\mathop\mathrm{arg max}} 

\def\reals{\mathbb R}

\def\Tcal{\mathcal T}

\def\thetabar{\overline{\theta}}

\newtheorem{theorem}{Theorem}
\newtheorem{lemma}{Lemma}
\newtheorem{proposition}{Proposition}

\newtheorem{remark}{Remark}

\usepackage[usenames,dvipsnames,table]{xcolor}

\usepackage{tikz, pgfplots}
\title{\LARGE \bf
Electric Vehicle Battery Sharing Game for Mobile Energy Storage Provision in Power Networks
}

\author{Utkarsha Agwan, Junjie Qin, Kameshwar Poolla, and Pravin Varaiya
\thanks{U. Agwan, K. Poolla and P. Varaiya are with the Department of Electrical Engineering and Computer Science, University of California at Berkeley
        {\tt\small \{uagwan, poolla, varaiya\}@berkeley.edu}}
\thanks{J. Qin is with the School of Electrical and Computer Engineering, Purdue University {\tt\small jq@purdue.edu}}%
}

\IEEEoverridecommandlockouts
\IEEEpubid{\text{\begin{minipage}[t]{\textwidth}\ \\[10pt]
       {A shorter version of this article has been accepted for presentation at the 2022 IEEE 61st Conference on Decision and Control (CDC)
        \copyright 2022 IEEE. Personal use of this material is permitted. Permission from IEEE must be obtained for all other uses, in any current or future media, including reprinting/republishing this material for advertising or promotional purposes, creating new collective works, for resale or redistribution to servers or lists, or reuse of any copyrighted component of this work in other works.}
\end{minipage}}}

\begin{document}

\maketitle


\begin{abstract}
Electric vehicles (EVs) equipped with a bidirectional charger can provide valuable grid services as mobile energy storage. However, proper financial incentives need to be in place to enlist EV drivers to provide services to the grid. In this paper, we consider two types of EV drivers who may be willing to provide mobile storage service using their EVs: commuters taking a fixed route, and on-demand EV drivers who receive incentives from a transportation network company (TNC) and are willing to take any route. We model the behavior of each type of driver using game theoretic methods, and characterize the Nash equilibrium (NE) of an EV battery sharing game where each EV driver withdraws power from the grid to charge its EV battery at the origin of a route, travels from the origin to the destination, and then discharges power back to the grid at the destination of the route. The driver earns a payoff that depends on the participation of other drivers and power network conditions. We characterize the NE in three situations: when there are only commuters, when there are only on-demand TNC drivers, and when the two groups of drivers co-exist. In particular, we show that the equilibrium outcome supports the social welfare in each of these three cases.
\end{abstract}

\def\Ds{\Delta^\mathrm{S}}
\def\Dm{\Delta^\mathrm{M}}
\def\eb{\mathbf{e}}

\section{Introduction}

Renewable generation is fast becoming the cheapest generation alternative, and scaling up variable renewable generation will require investment in electricity storage which can accommodate the uncertain and uncontrollable output. This trend is accompanied by a series of mandates in states like California and Massachusetts, which has spurred investment in utility scale battery projects \cite{GTMstorage20}. At the same time, there is a push to `electrify everything', i.e., move consumption from non-electric energy to electricity, which can then be supplied from clean, carbon-free resources like solar and wind. 

Transportation causes 29\% of global CO$_2$ emissions\cite{EPA-emissions}, and electrifying transport is an important step in any climate change mitigation plan. While transport sectors like long-haul trucking, shipping and aviation are hard to electrify, the passenger vehicle sector has seen rapid electrification in the last 10 years with 9\% of global car sales being electric vehicles (EVs) in 2021 \cite{IEA-EVoutlook}. EVs are equipped with batteries which can also be used to function as \rev{\emph{mobile energy storage}} in the power network with the help of bidirectional chargers, by charging at the origin of a route and discharging at the destination, thus moving energy across both space and time.  Such mobile energy storage in the form of EV batteries can help avoid time consuming and expensive transmission line upgrades, and also serve as energy storage on the grid \cite{agwan2021marginal}.  

However, unlike utility scale battery projects that can be dispatched by the power system operator, these EVs \rev{are owned and operated by individual drivers.} 
EV drivers will make independent decisions on whether to provide mobile storage service based on their individual costs and the value they create, which will be determined by the operation of the power network and the amount of mobile storage capacity available in the grid. They may have different motivations: some EV drivers may function as \emph{commuters}, i.e., travel along fixed routes, and some EV drivers could be available \emph{on-demand} to travel along specific routes to provide mobile storage service.
A fundamental question arises: \emph{will the market equilibrium lead to a socially desirable level of mobile storage capacity?}


This paper examines this question in three contexts: a) when there are a number of commuter EVs traveling along fixed routes in the power network which can provide mobile storage service along those routes, b) when there are a number of on-demand EV drivers which can provide mobile storage service along any route, and c) when there is a mix of commuter EVs and on-demand EVs in the network. The mobile storage service is provided in a wholesale market, and a transmission-constrained two-period economic dispatch problem is solved to determine the operation of the grid and mobile storage. This also determines the locational marginal prices, which are used to compensate mobile storage service providers. We make the following contributions to the literature: a) we develop novel game theoretic models in the context of sharing EV batteries as mobile energy storage, which incorporate both operation constraints of the power network and incentives for the EV drivers; 
b) we explicitly characterize the Nash Equilibrium (NE) of the proposed EV battery sharing games together with several benchmarks, and establish that NE support social welfare for all our settings.

Our work is built upon two lines of recent research. The first is game theoretic analysis of storage sharing. Among many papers in this area, the closest related works include:  \cite{kalathil2017sharing}, which studies the storage sharing and investment decisions of a collection of firms without considering network constraints, and \cite{qin2019distributed}, which analyzes the distributed storage investment game in power networks. \cite{carli2019distributed}  formulates a cooperative game for sharing energy storage within a residential microgrid,  \cite{yang2020selling} formulates a Stackelberg game to model the sharing of cloud energy storage, and \cite{zhao2019virtual} studies a two-stage problem of a central storage owner sharing virtualized sections of the storage capacity with multiple users. 
The second line of research is the growing literature on utilizing EVs as mobile energy storage to provide grid services. See \cite{qin2021mobile} for the cost-benefit analysis of the business model of sharing EVs to help commercial and industrial electricity users to reduce demand charges, \cite{agwan2021marginal} for the joint optimization of power network and a fleet of mobile storage units, and \cite{he2021utility} for a simulation study of the value of truck based mobile storage units in the California power grid.

\section{Model}\label{sec:model}

Consider a setting where EVs can provide mobile storage service in the power network.

\subsection{Power network}
We consider a power network with $n$ buses, where the buses are indexed by $i \in \mathcal N:=\{1, ..., n\}$. For simplicity, we consider a daily operation setting and divide the day into two time periods: an off-peak time period followed by a peak time period. We denote the time periods by $t\in \mathcal T:= \{1,2\}$.
For each  time period, we denote the generation and load over different buses in the power network by $\textbf{g}^{(t)} \in \mathbb{R}^n$ and  $\textbf{d}^{(t)} \in \mathbb{R}^n$, respectively. The convex generation cost function for the generator at bus $i$ and time $t$ is given by $C_{i,t}(\cdot)$, and the total generation cost at time $t$ is given by 
\begin{equation}
    C_t(\textbf{g}^{(t)}) = \sum_{i = 1}^n C_{i,t}(g_i^{(t)}), \quad t \in \mathcal T.
\end{equation}
The value of serving load at bus $i$ and time $t$ is given by the concave function $B_{i,t}(\cdot)$, which quantifies the utility derived by the consumers who use the supplied electric energy. 
Let the total consumer utility be
\begin{equation}
    B_t(\textbf{d}^{(t)}) = \sum_{i=1}^n B_{i,t}(d_i^{(t)}), \quad t \in \mathcal T.
\end{equation}
The buses in the power network are connected by transmission lines, and the power flow along each line should not exceed the line capacity. Additionally, the total power injection into the network at any time should be zero. We model these linearized AC power flow constraints by 
\begin{align}\label{eq:inj-constraints}
    \mathbf{1}^\top \textbf{p}^{(t)} = 0, \quad t \in \mathcal T, \\
    H \textbf{p}^{(t)} \leq \bar{f}, \: \quad t \in \mathcal T,
\end{align}
where $\textbf{p}^{(t)}\in \reals^n$ 
denotes the vector of power injections at each bus at time $t$. Here $H\in \reals^{m \times n}$ is the shift-factor matrix, where $m$ is the number of transmission constraints, and $\bar{f}$ denotes the line capacities.  


\subsection{EVs as mobile storage in the power network}
An EV can function as a mobile storage unit in the power network by charging at one bus during the off-peak period, moving to another bus and then discharging there during the peak period. In a power network with $n$ buses, there are a total of $n^2$ routes that the EV can take, which include the ``route'' where the EV stays at the same location. 

Consider an aggregate mobile storage capacity $S_{i,j}$ moving from bus $i$ at $t=1$ to bus $j$ at $t=2$, which comprises of all the EVs moving along that route and providing mobile storage service. The charging/discharging operation vector along route $i \rightarrow j$ is given by $\mathbf{u}_{i,j}\in \reals^2$, where positive values of $u_{i,j}^{(t)}$ indicate charging and negative values indicate discharging. We assume that each EV starts with an empty battery at $t=1$ and the aggregate state-of-charge at the end of time period $t\in \Tcal$ is given by $ \sum_{\tau = 1}^{t} u_{i,j}^{(\tau)}$, which is the sum of charging/discharging operations until that time. The state of charge must satisfy the energy capacity constraint of the aggregate storage capacity, i.e., 
\begin{align}
    0 \leq \textstyle\sum_{\tau = 1}^{t} u_{i,j}^{(\tau)} \leq S_{i,j}, \quad t \in \mathcal T.
\end{align}
Alternatively, we can write the constraint as 
\begin{align}
    \mathbf{0} \leq L \mathbf{u}_{i,j} \leq S_{i,j}\mathbf{1},
\end{align}
where $L\in \reals^{2\times 2}$ is a lower triangular matrix with $L_{t,t'}=1$ for all $t \ge t'$.
The vector of mobile storage capacities on all routes in the network is given by $\mathbf{S} \in \mathbb{R}^{n^2}$, and has an element $S_{i,j}$ corresponding to each route in the network. The power injection at each bus is the sum of generation and aggregate storage operation minus the demand, i.e.,
\thickmuskip=0mu
\begin{align}\label{eq:power-injection}
    p_i^{(1)} = g_i^{(1)} - d_i^{(1)} - \sum_{j=1}^n u_{i,j}^{(1)} ,   p_i^{(2)} = g_i^{(2)} - d_i^{(2)} - \sum_{i=1}^n u_{i,j}^{(2)}.
\end{align}
\thickmuskip=5mu plus 3mu minus 1mu
In the first time period, the storage operation $u_{i,j}^{(1)}$ occurs at the route origin, i.e., bus $i$, while in the second time period the operation $u_{i,j}^{(2)}$ occurs at the route destination, i.e., bus $j$. This explains the asymmetrical definition of power injection.


\subsection{Economic dispatch with mobile storage}
In a centralized optimization setting, the system operator dispatches generation and storage to supply the flexible load at each bus. Given the aggregate mobile storage capacity moving along each route, the system operator solves the multi-period economic dispatch problem to determine the operation of the grid and aggregate storage capacities:
\begin{subequations} \label{eq:ED}
	\begin{align} 
		J(\mathbf{S}) = \min_{\mathbf{p}, \mathbf{g}, \mathbf{d},\mathbf{u}} \quad & \sum_{t\in \mathcal T} C_t (\mathbf{g}^{(t)}) - B_t(\mathbf{d}^{(t)})\\
		\mbox{s.t.} \quad & p_i^{(1)} = g_i^{(1)} - d_i^{(1)} - \textstyle\sum_{j=1}^n u_{i,j}^{(1)} ,\label{eq:con:np1} \\
		& p_i^{(2)} = g_i^{(2)} - d_i^{(2)} - \textstyle\sum_{i=1}^n u_{i,j}^{(2)}, \label{eq:con:np2}\\
        & \mathbf{1}^\top \textbf{p}^{(t)} = 0, \quad  t  \in  \mathcal T, \label{eq:con:netzero}\\
   &  H \textbf{p}^{(t)} \leq \bar{f},   \quad t  \in  \mathcal T, \label{eq:con:line-cap}\\
		&   \mathbf{0} \le L \mathbf{u}_{i,j} \le S_{i,j} \mathbf{1},  \quad i,j  \in  \mathcal N. \label{eq:con:storage-cap}
	\end{align}
\end{subequations}
Denote the optimal dual variables associated with constraints \eqref{eq:con:np1} and \eqref{eq:con:np2} by $\boldsymbol{\lambda}^{(1)}, \boldsymbol{\lambda}^{(2)} \in \mathbb{R}^n$.
The locational marginal price (LMP) at bus $i$ at time $t$ is the marginal cost of generating an additional unit of energy, or the marginal value of supplying an additional unit of load. The dual variable $\lambda_i^{(t)}$ is the LMP at bus $i$ at time $t$.
The LMP at a load bus determines the payment made by loads, and the LMP at a generation bus is the price at which generators are compensated. Mobile storage must pay for the electricity it consumes through charging at the LMP, and is also compensated at the LMP for discharging. The LMP depends on the mobile storage capacity available, i.e., it is a function of $\mathbf{S}$ which is the vector of mobile storage capacities along each route in the network. We denote the LMPs by $\boldsymbol{\lambda}^{(t)}(\mathbf{S})$ to emphasize this dependence, but omit it in places for notational convenience.

\subsection{Commuter EV drivers with fixed routes}
Consider an EV \rev{driven by a commuter who regularly}  moves along one of the $n^2$ possible routes in the network. \rev{\emph{The route choice for individual drivers in this case is exogenous.}} The EV driver has the choice to use the EV battery as mobile storage along that route by charging at the origin and discharging at the destination. Each EV constitutes an infinitesimally small amount of storage traveling along a route, and we model the individual EVs as a continuum indexed by $k \in \rev{\mathcal{K}_{i,j}} = [0,1]$.
In providing this service, the EV driver buys electric energy at the locational marginal price (LMP) at the origin \rev{in the off-peak period}, and sells that energy at the destination LMP \rev{in the peak period}, thus capitalizing on the \rev{spatial-temporal} LMP difference along the route. The value gained by an EV driver moving along the route $i \rightarrow j$ by providing mobile storage service per unit of storage capacity is 
\begin{equation} \label{eq:lmp-diff}
    \lambda_j^{(2)}(\mathbf{S}) - \lambda_i^{(1)}(\mathbf{S}),
\end{equation}
where $\lambda_j^{(2)}(\mathbf{S})$ is the LMP at the destination node at time $2$ and $\lambda_i^{(1)}(\mathbf{S})$ is the LMP at the origin node at time $1$. 

In order to provide this service, the driver has to cycle through the EV battery capacity, thus causing some battery degradation. We model this battery degradation as a cost $\kappa$ of providing the service, which is incurred by the EV driver and is uniform across EV drivers. Further, the driver may have to park at a specialized charging station or wait longer than originally planned, and undergo some amount of inconvenience. 
\rev{For a driver $k\in \mathcal K_{i,j}$ traveling along the route $i \rightarrow j$,
we model this inconvenience as a cost $\theta_{k}$}. The \rev{collection of }drivers $\mathcal K_{i,j}$ have a range of inconvenience costs, which can be modeled as a continuous range of $\theta_k$ values. Since the commuter EV moves along the route in any case, the travel cost does not factor into the decision to provide mobile storage service. 

Each EV driver makes a decision on providing mobile storage service by comparing the value, i.e., the LMP difference in \eqref{eq:lmp-diff}, with the sum of battery degradation and inconvenience costs. We model this decision with a binary variable $s_k$, which is $1$ when the  \rev{driver $k$} provides mobile storage service, and $0$ when she does not. The payoff for  \rev{driver $k\in \mathcal K_{i,j}$} is
\begin{equation}\label{eq:k-payoff}
   \rev{ \pi_{k}(s_k, \mathbf{S}) }= \left[ \lambda_j^{(2)}(\mathbf{S}) - \lambda_i^{(1)}(\mathbf{S})  - \theta_{k} - \kappa \right] s_k
\end{equation}
per unit of storage capacity. 
The only difference in payoffs for drivers on the same route $i \rightarrow j $ is the inconvenience cost, which is different for each driver. Thus we can denote the decision to provide mobile storage service for each driver as a route-specific function $\sigma_{i,j} : \mathbb{R} \rightarrow \{0,1\}$ of the inconvenience cost, i.e.,
\begin{equation}
    s_k = \sigma_{i , j}(\theta_k),\quad  k \in \mathcal{K}_{i,j}, \quad i, j \in \mathcal N.
\end{equation}
The proportion of EVs that provide service is given by
\begin{equation}
    S_{i,j} = \mathbb{E} \sigma_{i,j}(\theta_k) = \int_{k \in \mathcal{K}_{i,j}} \sigma_{i,j}(\theta_k) \,\mathrm{d}\, F_{i,j} (\theta_k) \quad \leq 1,
\end{equation}
where $F_{i,j}(\cdot)$ is the cumulative distribution of inconvenience costs of the drivers on route $i \rightarrow j$. 

\begin{remark}
We can scale the actual storage capacity (in kWh), generation, and across the network such that the total mobile storage capacity available on each route ($\int_{k \in \mathcal{K}_{i,j}} d F (\theta_k)$) is scaled to be $1$. For routes with lower number of EVs, we can add `dummy' EVs with infinite inconvenience cost to obtain the same nominal number of EVs along each route. Non-linear coefficients of generation and load will need to be scaled appropriately.
\end{remark}


\subsection{On-demand EV drivers with flexible routes}

Consider an EV which is signed up with a transportation network company (TNC) and can be requisitioned to provide mobile storage service along any of the $n^2$ possible routes in the network. Each EV driver has the choice to use the EV battery as mobile energy storage by charging and then discharging along any of the possible routes, or to not provide the service at all. In a large fleet of EVs, each EV is an infinitesimally small amount of storage and we model the individual EVs as a continuum indexed by $\rev{\ell} \in \mathcal{L} = [0,1]$. 

The value for the EV driver moving along $i \rightarrow j$ is the same as that for an EV driver with a fixed route, given in \eqref{eq:lmp-diff}, and is dependent on the amount of mobile storage capacity in the network. However, the EV driver has to travel along $i \rightarrow j$ to provide this service, which she would not have otherwise since the sole purpose of the trip is providing mobile storage service. By providing this service, the EV battery will undergo some amount of degradation as well. The travel and battery degradation costs are modeled as a non-negative route-specific cost $\kappa_{i,j}$, and are the same for each EV traveling on this route. Further, the EV driver has to spend time and effort in traveling and providing mobile storage service and will need to be compensated for the inconvenience caused. This inconvenience can be modeled as a non-negative cost $\theta_\ell$ which is specific to driver $\ell$ but is route-independent.

Each EV driver compares the value and cost of providing mobile storage service to make a decision. An EV driver signed up with a TNC has $n^2 + 1$ possible choices: providing the service on any of the $n^2$ routes, or not providing the service at all. We denote the decision to provide service on route $i \rightarrow j $ with $\rev{s_{\ell;\, i,j}} \in \{0,1\} $, with $\sum_{i,j} \rev{s_{\ell;\, i,j}} \leq 1$. The payoff for the \rev{on demand EV driver $\ell \in \mathcal L$} is 
\begin{equation}\label{eq:l-payoff}
   \rev{ \pi_{\ell}}(\mathbf{s}_\ell, \mathbf{S}) = \sum_{i,j} \left( \lambda_j^{(2)}(\mathbf{S}) - \lambda_i^{(1)}(\mathbf{S}) - \theta_{\ell} - \kappa_{i,j} \right) \rev{s_{\ell;\, i,j}}
\end{equation}
per unit of storage capacity, \rev{where $\mathbf{s}_\ell = \{s_{\ell;\, i,j}\}_{i,j \in \mathcal N}\in \reals^{n^2}$}. The only difference in payoffs for different EVs is their inconvenience cost, which in turn determines their route choice. We can then denote the optimal service provision choice by $\mathbf{s}_{\ell} = \delta(\theta_\ell), \ell \in \mathcal{L}$, \rev{where $\delta: \reals \mapsto \{0,1\}^{n^2}$.}
We also define $s_\ell = \mathbf{1}^\top \mathbf{s}_\ell = \sum_{i,j} s_{\ell;i, j} \in \{0,1\}$ which denotes whether the EV provides service along any route in the network. If $s_\ell = 0$,  EV $\ell$ does not provide service along any route. We define $S_{i,j}$ as the amount of mobile storage available on $i \rightarrow j$. We have
\begin{equation}
S_{i,j}  = \int_{\ell \in \mathcal{L}} \delta(\theta_\ell)_{i,j} \,\mathrm{d}\, F(\theta_\ell) \le 1,
\end{equation}
where $\delta(\theta_\ell)_{i,j} = \rev{s_{\ell;\, i,j}}$ is the decision to provide service on route $i \rightarrow j$, and $F(\cdot)$ is the cumulative distribution of the inconvenience costs of the on-demand EVs. Note that $\sum_{i,j} S_{i,j} \le 1$ as well.

\subsection{Solution concepts}\label{sec:soln-concepts}

Both commuter and on-demand EVs can be operated by a variety of centralized operators and decentralized agents with different objectives. We begin by considering \rev{two} benchmark solution concepts:
\begin{enumerate}
    \item \emph{Myopic EV drivers:} EVs act in a decentralized manner and maximize their own individual payoffs without considering the effect of mobile storage service on the LMPs. They optimize their operation under the assumption that $\mathbf{S} = \mathbf{0}$.
    

    \item \emph{Social welfare maximizing operator:} A central operator optimizes the mobile storage service provision of all the EVs to maximize social welfare, which is defined as \rev{the surplus received by both the EVs and the electricity market participants, including the generators and load.}
    
\end{enumerate}
\rev{The precise mathematical description of these benchmarks differs depending on the type of EV drivers under consideration (commuters or on-demand drivers), and will be provided in subsequent sections.}

We then consider the operation of EVs which operate in a decentralized manner to optimize their individual payoffs, thus participating in an EV battery sharing game. We define three game settings: 
\begin{enumerate}
    \item \emph{Commuter EVs only:}
    The set of players  is $\cup_{i,j \in \mathcal N} \mathcal K_{i,j}$, where each player has a decision of whether to provide mobile storage service or not. The payoff of player $k \in \mathcal K_{i,j}$ is defined as \eqref{eq:k-payoff}.

    \item \emph{On-demand EVs only:} 
    The set of players is $\mathcal L$, where each player chooses from $n^2$ routes to provide the mobile storage service or not to provide the service. The payoff of player $\ell \in \mathcal L$ is defined as \eqref{eq:l-payoff}.
    
    \item \emph{Both commuter and on-demand EVs:} In this case, both types of  players coexist, with their decisions and payoff functions defined as before. 
\end{enumerate}
\rev{We utilize \emph{Nash Equilibrium} (NE) as the solution concept, under which no player has incentive to unilaterally change its decision.}
As the game is an aggregate game, i.e., each player's action only impact others' payoffs via the aggregate storage capacities, we will refer to the aggregate storage capacities induced by a NE as NE storage capacities.
For each setting, we will  compare the NE  to the benchmarks discussed previously.

\section{Commuter EVs: Fixed Routes}\label{sec:fixed}

\rev{In this section, we consider the setting where there are only commuter EVs providing  mobile storage services to the grid, and characterize the market driven equilibrium outcome for the EV battery sharing game. }
Each route in the power network has a population of EV drivers $k \in \mathcal K_{i,j}$ characterized by their inconvenience cost $\theta_k$, which are otherwise interchangeable. In order to define the optimal mobile storage service for each of the solution concepts, we partition the population of EVs on route $i \rightarrow j$ into $\mathcal{K}_{i,j}^{+}$ and $\mathcal{K}_{i,j}^{-}$ for each situation, where EVs in $\mathcal{K}_{i,j}^{+}$ provide mobile storage service, and EVs in $\mathcal{K}_{i,j}^{-}$ do not. We posit that the EVs in $\mathcal{K}_{i,j}^{+}$ necessarily have a lower inconvenience cost than the EVs in $\mathcal{K}_{i,j}^{-}$ for each solution concept discussed in section~\ref{sec:soln-concepts} (which will be mathematically defined subsequently), i.e.,



\begin{proposition}\label{prop:theta-order}
For each solution concept discussed in this paper, there exists a threshold $\thetabar_{i,j}$ such that
\begin{subequations}
    \begin{align}
        \mathcal{K}_{i,j}^{+} = \{k\in \mathcal K_{i,j}: \theta_k \le \thetabar_{i,j}\}, \quad i,j \in \mathcal N, \\
         \mathcal{K}_{i,j}^{-} = \{k\in \mathcal K_{i,j}: \theta_k > \thetabar_{i,j}\},  \quad i,j \in \mathcal N.
    \end{align}
\end{subequations}
\end{proposition}
\begin{proof}
We split the discussion into two cases:
\begin{enumerate}
    \item \emph{Independent decisions by EVs:} Each EV makes a decision to provide mobile storage service based only on its payoff. The payoffs, and consequently the service decisions for two different EVs are only differentiated by their inconvenience cost $\theta_k$. An EV with a higher $\theta_k$ will necessarily have a lower payoff \eqref{eq:k-payoff}, and if this EV decides to provide service, then any EV with a higher payoff (i.e., lower $\theta_k$) will also provide service. 
    \item \emph{Centrally controlled EVs:} From the perspective of the power system operator, EVs are undifferentiated except by their inconvenience costs. The system operator will preferentially choose EVs with lower inconvenience costs instead of those with higher inconvenience costs in order to minimize total social cost. 
\end{enumerate}
In both cases, there will be a marginal EV on route $i \rightarrow j$ with inconvenience cost $\thetabar_{i,j}$ such that all EVs on that route with lower $\theta$ will provide service, and those with higher $\theta$ will not. 
\end{proof}

\subsection{Benchmarks}

We now characterize the operation in the benchmarks discussed in Section \ref{sec:soln-concepts} \rev{when there are only commuters}. 

\subsubsection{Myopic EV drivers}
Each myopic EV owner traveling along $i \rightarrow j$ maximizes $\pi_{k}(s_k, \mathbf{0})$.
The optimal decision would be to set
\begin{align}
    s_k^{\text{myop}} = \begin{cases}
    1, & \text{if} \: \lambda_j^{(2)}(\mathbf{0}) - \lambda_i^{(1)}(\mathbf{0}) - \theta_k - \kappa \geq 0, \\
    0, & \text{otherwise,}
    \end{cases}
\end{align}
which gives us threshold inconvenience cost for each route
\begin{equation}
    \thetabar_{i,j}^{\text{myop}} =  \lambda_j^{(2)}( \mathbf{0}) - \lambda_i^{(1)}(\mathbf{0}) - \kappa ,
\end{equation}
and the mobile storage proportion $S^{\text{myop}}_{i,j} = F_{i,j}( \thetabar_{i,j}^{\text{myop}})$. 

\subsubsection{Social welfare maximizing operator}
A social welfare maximizing operator solves the following problem:
\begin{subequations}\label{eq:sw}
\begin{align}
    \min_{\mathbf{S}, \boldsymbol{\thetabar}} & \quad J(\mathbf{S}) + \sum_{i,j}  \int_{\theta_k \leq  \thetabar_{i,j}} (\theta_k + \kappa) \,\mathrm{d}\,F(\theta_k), \\
   \text{s.t.} & \quad S_{i,j} = F_{i,j} (\thetabar_{i,j}),\quad i,j \in \mathcal N, \label{eq:con-sw-fix}
\end{align}
\end{subequations}
where $J(\mathbf{S})$ is the optimal solution of the economic dispatch problem in \eqref{eq:ED}. The social cost is taken to be the sum of generation cost, inconvenience and battery degradation costs for the EVs, minus the value of supplying electricity to loads. Then we have
\begin{lemma}\label{lem:sw-fix}
The inconvenience cost threshold $\thetabar_{i,j}^{\text{sw}}$ and the corresponding aggregate storage capacity $S_{i,j}^{\text{sw}}$, $i,j\in \mathcal N$, for the socially optimal operation  is given by \rev{the solution of}
\begin{subequations}\label{eq:theta-sw}
\noeqref{eq:1,eq:2}
    \begin{align}
  & \thetabar_{i,j}^{\text{sw}} = \lambda_j^{(2)}(\mathbf{S}^{\text{sw}}) - \lambda_i^{(1)}(\mathbf{S}^{\text{sw}}) - \kappa, &\quad i, j \in \mathcal N. \label{eq:1} \\
 & \rev{ S_{i,j}^{\text{sw}} = F_{i,j} (\thetabar_{i,j}^{\text{sw}})}, &\quad i,j \in \mathcal N. \label{eq:2}
\end{align}
\end{subequations}

\end{lemma}

\begin{proof}
We can eliminate one variable in \eqref{eq:sw} by enforcing the equality constraint \eqref{eq:con-sw-fix}. We set the gradient of the objective in the unconstrained problem to zero, i.e.
\begin{align}
    S^{\text{sw}}_{i,j} : \nabla_{S_{i,j}} J(\mathbf{S}) |_{\mathbf{S}^{\text{sw}}}+   (\thetabar_{i,j}^{\text{sw}} + \kappa ) = 0.
\end{align}
From \cite{agwan2021marginal}, we know that 
\begin{equation}
    \nabla_{S_{i,j}} J(\mathbf{S}) = - (\lambda_j^{(2)}(\mathbf{S}) - \lambda_i^{(1)}(\mathbf{S}))_+,
\end{equation}
and we can ignore the positive part operator since a non-zero $S_{i,j}^{\text{sw}}$ will necessitate a non-negative $\thetabar_{i,j}^{\text{sw}}$, which ensures that $\lambda_j^{(2)}(\mathbf{S}) - \lambda_i^{(1)}(\mathbf{S}) \geq 0$, and $\kappa$ is necessarily non-negative. 
\end{proof}

The marginal increase in social welfare on increasing mobile storage along route $i \rightarrow j$ is given by the decrease in power system cost less the inconvenience and battery degradation costs incurred by the commuter EV, i.e., 
\begin{equation}\label{eq:sw-inc-fixed}
    (\lambda_j^{(2)}(\mathbf{S}) - \lambda_i^{(1)}(\mathbf{S}))_+ - \theta_{i,j} - \kappa.
\end{equation}


\subsection{Nash equilibrium}
Consider a situation where all the EVs are owned and operated by distributed entities, e.g., the case where they are all personal vehicles used for transport, and each EV driver participates in the EV battery sharing game independently.
We can classify EVs into two classes: those which are providing mobile storage service at equilibrium and those which are not.
At the equilibrium, no EV will be better off switching from one class to another. 

At the NE, given the aggregate storage capacities, each EV maximizes its payoff, i.e., 
\begin{equation}
    \max_{s_k} \quad  \pi_{k}(s_k, \mathbf{S}) .
\end{equation}
If each EV has a small storage capacity, then the operational decision of one EV does not impact the LMPs, and
\begin{equation}\label{eq:s_k-NE}
    s_k^{\text{NE}} = \sigma^{\text{NE}}_{i,j}(\theta_k) = \begin{cases}
    1, \quad & \text{if} \: 
    \pi_{k}(1, \mathbf{S}^{\text{NE}})\ge 0,\\
    0, \quad & \text{otherwise},
    \end{cases}
\end{equation}
\rev{for $k \in \mathcal K_{i,j}$}.
In other words, any EV which can obtain a non-negative payoff decides to provide mobile storage service. 
\begin{lemma}\label{lem:thetabar-ne}
The Nash equilibrium inconvenience cost threshold $\thetabar^{\text{NE}}_{i,j}$ and the corresponding aggregate storage capacity $S^{\text{NE}}_{i,j}$, $i, j \in \mathcal{N}$ are given by the solution of
\begin{subequations}\label{eq:theta-NE}
\noeqref{eq:1,eq:2}
    \begin{align}
        & \thetabar^{\text{NE}}_{i,j}  =  \lambda_j^{(2)}(\mathbf{S}^{\text{NE}}) - \lambda_i^{(1)}(\mathbf{S}^{\text{NE}})  - \kappa, & \quad i,j \in \mathcal{N}. \label{eq:1}\\
        & S^{\text{NE}}_{i,j} = F_{i,j}(\thetabar^{\text{NE}}_{i,j}) , & \quad i,j \in \mathcal{N}. \label{eq:2}
    \end{align}
\end{subequations}

\end{lemma}
\begin{proof}
Consider an EV with $\theta > \thetabar^{\text{NE}}_{i,j}$. We posit that at equilibrium, this EV does not provide mobile storage service. The payoff for this EV is given by 
\begin{align*}
& \lambda_j^{(2)}(\mathbf{S}^{\text{NE}}) - \lambda_i^{(1)}(\mathbf{S}^{\text{NE}})  - \kappa - \theta  \\
< \: & \lambda_j^{(2)}(\mathbf{S}^{\text{NE}}) - \lambda_i^{(1)}(\mathbf{S}^{\text{NE}})  - \kappa - \thetabar^{\text{NE}}_{i,j} = 0,
\end{align*}
i.e., the payoff is negative, and the EV has no incentive to deviate from the equilibrium decision and start to provide mobile storage service. 

Next, consider the complementary case, i.e., an EV with $\theta \leq \thetabar^{\text{NE}}_{i,j}$. We posit that at equilibrium, this EV will provide mobile storage service. The payoff for this EV is given by 
\begin{align*}
& \lambda_j^{(2)}(\mathbf{S}^{\text{NE}}) - \lambda_i^{(1)}(\mathbf{S}^{\text{NE}})  - \kappa - \theta  \\ 
\geq \: & \lambda_j^{(2)}(\mathbf{S}^{\text{NE}}) - \lambda_i^{(1)}(\mathbf{S}^{\text{NE}})  - \kappa - \thetabar^{\text{NE}}_{i,j} = 0,
\end{align*}
i.e., is non-negative, and the EV has no incentive to deviate from the equilibrium decision and stop providing service. 
\end{proof}


We can relate the equilibrium mobile storage service with the socially optimal solution as:
\begin{theorem}
Any aggregate storage capacity corresponding to the Nash equilibrium for commuter EVs supports the social welfare.
\end{theorem}
\begin{proof}
There is a one-to-one correspondence of the socially optimal aggregate storage capacity given in \eqref{eq:theta-sw} and the Nash equilibrium aggregate storage given in \eqref{eq:theta-NE}.
\end{proof}

\subsection{Example}
\begin{figure} 
\centering
\includegraphics[width=.42\textwidth]{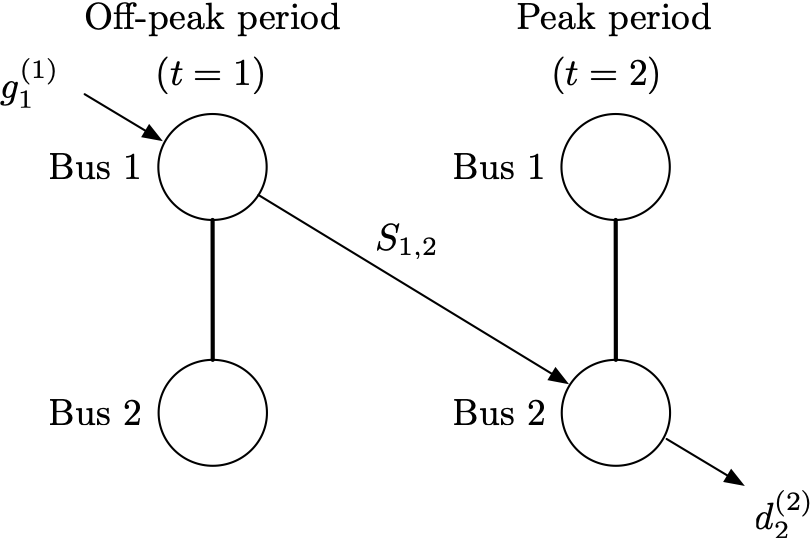}
\caption{Two bus network }\label{fig:2bus}
\end{figure}

Our results on the social welfare maximizing solution and NE depend on solving a system of nonlinear equations. To gain explicit analytical insight, we consider a simple example with two period and two buses shown in Fig. \ref{fig:2bus}.  
For the network, bus 2 has a load at $t=2$, and bus 1 has a generator. There is some mobile storage capacity ($S_{1,2}$) which moves from bus 1 at $t=1$ to bus 2 at $t=2$. The generation cost is given by $C_t(g) = ag^2 + bg, t \in \{1,2\}$ and the value of supplying load is given by $B_t(d) = c d , t \in \{1,2\}$.
The LMPs for this network are given by
\begin{align}
    \lambda_1^{(1)} = 2a\min \left\{ S_{1,2},\frac{c-b}{2a} \right\} + b, \quad     \lambda_2^{(2)} = c. 
\end{align}
The optimal mobile storage levels for each of the benchmarks for this network is given by 
\begin{enumerate}

    \item \emph{Myopic EV drivers:} The LMPs when $\mathbf{S} = \mathbf{0}$ are $\lambda_1^{(1)} = b$, $\lambda_2^{(2)} = c$, and we have 
    \begin{equation}
        S^{\text{myop}}_{1,2} = F_{1,2}(\thetabar_{1,2}^{\text{myop}} ) = F_{1,2} (c - b- \kappa) .
    \end{equation}

    \item \emph{Social welfare maximizing operator:} The optimal decision is given by the solution of the fixed point equation
    \begin{align}
        & \quad S^{\text{sw}}_{1,2}  = F_{1,2}(c - 2aS^{\text{sw}}_{1,2} -b - \kappa),
    \end{align}
\end{enumerate}
On comparing these values, we find 
\begin{equation}
    S^{\text{myop}}_{1,2} \geq S^{\text{sw}}_{1,2} = S^{\text{NE}}_{1,2} .
\end{equation}
The intuition behind this is that myopic EV drivers tend to over-commit to providing mobile storage service, since they do not factor the reduction of the LMP difference into their decision. 

\section{On-Demand EVs: Flexible Routes}\label{sec:flex}
In this section, we consider the setting where there are only on-demand EVs providing mobile storage services to the grid, and characterize the market driven equilibrium outcome for the EV battery sharing game. 
The network has a population of on-demand EVs $\ell \in \mathcal L$, which are characterized by their inconvenience cost $\theta_\ell$, and are otherwise interchangeable. 
In order to define the optimal storage service, we partition the network-wide population of EVs into $\mathcal{L}^+$ and $\mathcal{L}^-$ for each of the solution concepts,
where EVs in $\mathcal{L}^+$ provide mobile storage service on any one route in the power network and EVs in $\mathcal{L}^-$ do not provide mobile storage on any route. We can extend Proposition \ref{prop:theta-order} and define a network-wide inconvenience cost threshold $\thetabar$ for each solution concept, such that 
\begin{subequations}
    \begin{align}
        \mathcal{L}^{+} = \{l\in \mathcal L: \theta_{\ell} \le \thetabar\}, \\
        \mathcal{L}^{-} = \{l\in \mathcal L: \theta_{\ell} > \thetabar\}.
    \end{align}
\end{subequations}


\subsection{Benchmarks}

We now characterize the operation of on-demand EVs in some of the benchmarks discussed in Section \ref{sec:soln-concepts}.


\subsubsection{Myopic EV drivers}
The myopic EV driver indexed by $\ell \in \mathcal{L}$ chooses $\mathbf{s}_\ell$ to maximize $\pi_{\ell}(\mathbf{s}_\ell, \mathbf{0})$. Since the only difference in payoffs for EVs is the inconvenience cost $\theta_\ell$, the route with the maximum potential payoff
will attract all the myopic EV drivers. Let this route be  $i^* \rightarrow j^*$, where
\begin{equation}\label{eq:flex-myop-route}
    (i^*, j^*) = \argmax_{i,j}\,\,  \lambda_j^{(2)}(\mathbf{0}) - \lambda_i^{(1)}(\mathbf{0})- \kappa_{i,j}.
\end{equation}
Then the  decision of driver $\ell \in \mathcal L$ to provide service is 
\begin{equation}
    s_{\ell;i^*,j^*}^{\text{myop}} = \begin{cases}
    1, \quad & \text{if} \:  \lambda_{j^*}^{(2)}(\mathbf{0}) - \lambda_{i^*}^{(1)}(\mathbf{0})- \kappa_{i^*,j^*} \geq \theta_\ell, \\
    0, \quad & \text{otherwise},
    \end{cases}
\end{equation}
and with $s_{\ell;\, i,j}^{\text{myop}} =  0 $ for $ (i, j) \neq (i^*, j^*)$. This gives us the inconvenience cost threshold 
\begin{equation}
    \thetabar^{\text{myop}} = \lambda_{j^*}^{(2)}(\mathbf{0}) - \lambda_{i^*}^{(1)}(\mathbf{0})- \kappa_{i^*,j^*},
\end{equation}
where $i^*, j^*$ are defined as in \eqref{eq:flex-myop-route}. The total storage capacity for route $(i^*, j^*)$ is given by $S_{i^*, j^*}^{\text{myop}} = F(\thetabar^{\text{myop}})$,
and $S_{i,j}^{\text{myop}}=0$ for all other routes.


\subsubsection{Social welfare maximizing operator}
A central operator that maximizes social welfare (or equivalently minimizing the social cost) solves the following problem:
\begin{subequations}\label{eq:sw-flex}
\begin{align}
    \min_{\mathbf{S}, \thetabar} \quad & J(\mathbf{S}) + \sum_{i,j} \kappa_{i,j}S_{i,j} + \int_{\theta_\ell \leq \thetabar}  \theta_\ell \,\,\mathrm{d}\, F(\theta_\ell)\\
    \text{s.t.} \quad & \mathbf{S} \geq \mathbf{0}, \: \mathbf{1}^\top \mathbf{S} = F(\thetabar),
\end{align}
\end{subequations}
where $J(\mathbf{S})$ is the optimal cost of the economic dispatch problem in \eqref{eq:ED}. The social cost is taken to be the sum of generation cost, inconvenience, travel and battery degradation costs for the EVs, minus the value of supplying electricity to loads.  


We can define the storage capacity on each route by $S^{\text{sw}}_{i,j}$, and we know that $\sum_{i,j} S^{\text{sw}}_{i,j} = F(\bar{\theta}^{\text{sw}})$, where $\bar{\theta}^{\text{sw}}$ is the network-wide threshold of inconvenience costs determined by solving \eqref{eq:sw-flex}. From \cite{agwan2021marginal} we know that the value of increasing mobile storage capacity is given by
\begin{equation}
    - \nabla_{S_{i,j}} J(\mathbf{S}) = (\lambda_j^{(2)}(\mathbf{S}) - \lambda_i^{(1)}(\mathbf{S}))_+,
\end{equation}
which is the LMP increase along the route. However, increasing storage capacity along a route also increases the travel and inconvenience costs that need to be paid. The operator will add mobile storage capacity which maximizes the increase in social welfare
\begin{equation} \label{eq:inc-sw-flex}
    \left( (\lambda_j^{(2)}(\mathbf{S}) - \lambda_i^{(1)}(\mathbf{S}))_+ - \theta_\ell - \kappa_{i,j} \right)_+.
\end{equation}
 We can ignore the inner positive part operator in this equation when we formulate our storage operation decision, since the LMP difference will necessarily be non-negative for the entire expression to be non-negative. 
 The socially optimal storage operation of on-demand EVs can be formulated as a route choice 
 \begin{equation}\label{eq:flex-sw-route}
    (i^*, j^*) = \argmax_{i,j}  \lambda_j^{(2)}(\mathbf{S}^{\text{sw}}) - \lambda_i^{(1)}(\mathbf{S}^{\text{sw}}) - \kappa_{i,j} ,
\end{equation}
 and a mobile storage service provision choice given by
\medmuskip=2mu
\thinmuskip=2mu
\thickmuskip=3mu
 \begin{subequations}\label{eq:s_ell-sw}
\begin{align}
& s_{\ell;\, i,j}^{\text{sw}} = 0, \quad \text{if} \: (i,j) \neq (i^*, j^*), \\
   & s_{\ell;i^*,j^*}^{\text{sw}} = \begin{cases}
    1, \: \text{if} \:  \lambda_{j^*}^{(2)}(\mathbf{S}^{\text{sw}}) - \lambda_{i^*}^{(1)}(\mathbf{S}^{\text{sw}}) - \theta_\ell - \kappa_{i^*,j^*} \geq 0,\\
    0, \: \text{otherwise},
    \end{cases}  \label{eq:s_ell-sw-inc-only} 
\end{align}
\end{subequations}
\medmuskip=4mu
\thinmuskip=3mu
\thickmuskip=5mu
which ensures that storage is only added if it increases the social welfare. 

\begin{lemma}\label{lem:sw-flex}
The network-wide inconvenience cost threshold for socially optimal operation of on-demand EVs is the solution of
\begin{subequations} \label{eq:thetabar-sw-flex}
\noeqref{eq:1,eq:2}
    \begin{align}
        & \thetabar^{\text{sw}} = \lambda_{j^*}^{(2)}(\mathbf{S}^{\text{sw}}) - \lambda_{i^*}^{(1)}(\mathbf{S}^{\text{sw}})  - \kappa_{i^*,j^*}, \label{eq:1} \\
        & \mathbf{1}^\top \mathbf{S}^{\text{sw}} = F(\thetabar^{\text{sw}}), \label{eq:2}
    \end{align}
\end{subequations}
where $i^*, j^*$ are defined as in \eqref{eq:flex-sw-route}.
\end{lemma}

\begin{proof}
\begin{enumerate}
\item First, consider $\mathcal{L}^{-}$ is non-empty, i.e., there are some EVs which do not provide mobile storage along any route at the socially optimal solution. This means that the marginal increase in social welfare on adding an EV to any route in the network is non-positive, since otherwise the unutilized mobile storage would be dispatched along some route in the network. There are two types of routes in the network:

\begin{enumerate}
        \item Route $i \rightarrow j$ has some non-zero mobile storage capacity, i.e., $S_{i,j}^{\text{sw}} > 0$. The marginal increase in social welfare on adding an additional EV to this route is zero, and
        \begin{align}
    \lambda_j^{(2)}(\mathbf{S}^{\text{sw}}) - \lambda_i^{(1)}(\mathbf{S}^{\text{sw}}) - \thetabar^{\text{sw}} - \kappa_{i,j}  = 0.
\end{align}
We can ignore the positive part operator in \eqref{eq:inc-sw-flex} since $\thetabar^{\text{sw}}, \kappa_{i,j}$ are necessarily non-negative. 
    
        \item Route $i' \rightarrow j'$ has zero mobile storage capacity, i.e., $S_{i',j'}^{\text{sw}} = 0$. Adding an EV to this route causes a decline in social welfare, i.e.
    \begin{equation}
        \lambda_{j'}^{(2)}(\mathbf{S}^{\text{sw}}) - \lambda_{i'}^{(1)}(\mathbf{S}^{\text{sw}}) - \thetabar^{\text{sw}} - \kappa_{i',j'} \leq 0. 
    \end{equation}
  \end{enumerate} 

    \item Second, consider $\mathcal{L}^{-}$ is empty, then $\sum_{i,j} S^{\text{sw}}_{i,j} = \mathbf{1}^\top \mathbf{S}^{\text{sw}} = 1$ and all EVs are deployed along one of the routes in the network. The network wide inconvenience cost threshold is $\thetabar^{\text{sw}} \geq \max_l \:  \{ \theta_l \}$, and at least one route has a non-negative payoff, i.e.,
    \medmuskip=2mu
\thinmuskip=2mu
\thickmuskip=3mu
     \begin{align}
    \lambda_j^{(2)}(\mathbf{S}^{\text{sw}}) - \lambda_i^{(1)}(\mathbf{S}^{\text{sw}}) - \thetabar^{\text{sw}} - \kappa_{i,j}  \geq  0  \: \text{ for some } i,j.
\end{align}
\medmuskip=4mu
\thinmuskip=3mu
\thickmuskip=5mu
This indicates that if there were more EVs available with an inconvenience cost $\thetabar^{\text{sw}}$, it would be beneficial for them to provide mobile storage service.
\end{enumerate}
\end{proof}

\subsection{Nash equilibrium}
Consider a situation where all EVs are owned and operated by distributed agents, e.g., when they are owned by individuals who sign up on TNC platforms to earn money for providing mobile storage service. Each EV driver participates in an EV battery sharing game, and makes an independent decision on whether to provide mobile storage service and which route to provide it on based on the payoff $\pi_{\ell}(\mathbf{s}_\ell, \mathbf{S})$. 
At the equilibrium, EVs will provide service in a manner where no EV has an incentive to deviate from its chosen route and operation. 

At the NE, given the aggregate storage capacities, each EV maximizes its own payoff, i.e., chooses a route and operational decision according to 
\begin{subequations}
    \begin{align}
        \max_{\mathbf{s}_\ell} & \quad \pi_{\ell}(\mathbf{s}_\ell, \mathbf{S}) \\
        \text{s.t.} & \quad \mathbf{1}^\top \mathbf{s}_\ell \leq 1, 
    \end{align}
\end{subequations}
where $\mathbf{s}_\ell$ is the decision vector of $\rev{s_{\ell;\, i,j}}$ for all routes on the network for driver $\ell$. If each individual EV has a small storage capacity, then its operation decision will not impact LMPs and we can denote the optimal route choice of the marginal EV by 
\begin{equation}\label{eq:flex-route-ne}
    (i^*, j^* )= \argmax_{i,j}  \lambda_j^{(2)}(\mathbf{S}^{\text{NE}}) - \lambda_i^{(1)}(\mathbf{S}^{\text{NE}}) - \kappa_{i,j} ,
\end{equation}
 and the mobile storage service provision by
 \medmuskip=2mu
\thinmuskip=2mu
\thickmuskip=2mu
\begin{subequations}\label{eq:s_ell-ne}
\begin{align}
& s_{\ell;\, i,j}^{\text{NE}} = 0, \quad \text{if} \: (i,j) \neq (i^*, j^*), \\
   & s_{\ell;i^*,j^*}^{\text{NE}} = \begin{cases}
    1, \: \text{if} \:  \lambda_{j^*}^{(2)}(\mathbf{S}^{\text{NE}}) - \lambda_{i^*}^{(1)}(\mathbf{S}^{\text{NE}}) - \theta_\ell - \kappa_{i^*,j^*} \geq 0, \\
    0, \: \text{otherwise}.
    \end{cases} 
\end{align}
\end{subequations}
\medmuskip=4mu
\thinmuskip=3mu
\thickmuskip=5mu
For the equilibrium mobile storage $\mathbf{S}^{\text{NE}}$, there are two dimensions: each EV which provides mobile storage service on route $i \rightarrow j$ at equilibrium will be no better off if (a) it decides to stop providing the service, or (b) if it switches to a different route. Additionally, each EV which does not provide mobile storage service will be no better off if it does so on any route in the network. 
\begin{lemma}\label{lem:on-demand-ne}
The network-wide inconvenience cost threshold at equilibrium $\thetabar^{\text{NE}}$ and the corresponding aggregate storage capacity are given by the solution of
\begin{subequations}\label{eq:thetabar-ne-flex}
\noeqref{eq:1, eq:2}
    \begin{align}
        & \thetabar^{\text{NE}} = \lambda_{j^*}^{(2)}(\mathbf{S}^{\text{NE}}) - \lambda_{i^*}^{(1)}(\mathbf{S}^{\text{NE}})  - \kappa_{i^*,j^*}, \label{eq:1} \\
        & \mathbf{1}^\top \mathbf{S}^{\text{NE}} = F(\thetabar^{\text{NE}}), \label{eq:2}
    \end{align}
\end{subequations}
where $i^*, j^*$ are defined as in \eqref{eq:flex-route-ne}.
\end{lemma}


The Nash equilibrium operation of on-demand EVs providing mobile storage service is characterized through the following exhaustive list of cases, which can also serve as a proof of Lemma \ref{lem:on-demand-ne} by utilizing arguments similar to the ones in the proof for Lemma \ref{lem:thetabar-ne}:
\begin{enumerate}
    \item First, consider the case where $\mathcal{L}^-$ is not empty, i.e. there are some EVs which do not provide mobile storage service. Then, the marginal payoff for an additional EV on any route should be non-positive. There are two types of routes in the network:

    \begin{enumerate}
        \item Route $i \rightarrow j$ has some non-zero mobile storage capacity, i.e. $S_{i,j} > 0$. At equilibrium, this route provides zero payoff and there is no incentive for an EV $\ell \in \mathcal{L}^-$ to provide service on this route, and
        \begin{align}
        \thetabar^{\text{NE}} =     \lambda_j^{(2)}(\mathbf{S}^{\text{NE}}) - \lambda_i^{(1)}(\mathbf{S}^{\text{NE}})   - \kappa_{i,j} .
\end{align}

        \item Route $i' \rightarrow j'$ has zero mobile storage capacity, i.e. $S_{i',j'} = 0$. This route has a non-positive payoff, which is why no EV choses to provide service on that route, and
    \begin{equation}
        \lambda_{j'}^{(2)}(\mathbf{S}^{\text{NE}}) - \lambda_{i'}^{(1)}(\mathbf{S}^{\text{NE}}) - \thetabar^{\text{NE}} - \kappa_{i',j'} < 0. 
    \end{equation}
  \end{enumerate}

\item Second, consider the case where $\mathcal{L}^-$ is empty, i.e. all the EVs available provide mobile storage service on one route or the other. Then we have
\begin{subequations}
    \begin{align}
       \thetabar^{\text{NE}} & =  \lambda_{j_1}^{(2)}(\mathbf{S}^{\text{NE}}) - \lambda_{i_1}^{(1)}(\mathbf{S}^{\text{NE}}) - \kappa_{i_1,j_1}  \\
        & =   \lambda_{j_2}^{(2)}(\mathbf{S}^{\text{NE}}) - \lambda_{i_2}^{(1)}(\mathbf{S}^{\text{NE}})  - \kappa_{i_2,j_2}  \\
        & \geq  \lambda_{j_3}^{(2)}(\mathbf{S}^{\text{NE}}) - \lambda_{i_3}^{(1)}(\mathbf{S}^{\text{NE}})  - \kappa_{i_3,j_3}
    \end{align}
    \end{subequations}
    for all $i_1,j_1,i_2,j_2, i_3, j_3 \in \mathcal N$ where $S_{i_1,j_1}^{\text{NE}} > 0, S_{i_2,j_2}^{\text{NE}} > 0$ and $S_{i_3,j_3}^{\text{NE}} = 0$, i.e., the marginal payoff for routes with non-zero mobile storage capacity is the same throughout the network, and is higher than the marginal payoff for routes with zero mobile storage capacity at equilibrium. 
    This indicates that if there were more EVs available with an inconvenience cost $\thetabar^{\text{NE}}$, it would be profitable for them to provide mobile storage service. 
\end{enumerate}
We can relate the equilibrium mobile storage service to the socially optimal solution as:
\begin{theorem}
Any inconvenience cost threshold $\thetabar$ corresponding to the Nash equilibrium for on-demand EVs supports the social welfare.
\end{theorem}
\begin{proof}
There is a one-to-one correspondence of the equilibrium inconvenience cost threshold in \eqref{eq:thetabar-ne-flex} and the socially optimal inconvenience cost threshold in \eqref{eq:thetabar-sw-flex}.
\end{proof}

\section{Hybrid: Commuter and On-Demand EVs}\label{sec:both}

In this section, we consider the setting where there is a mix of commuter and on-demand EVs providing mobile storage services to the grid, and characterize the market driven equilibrium outcome as an EV battery sharing game. There is a population of commuter EVs on each route characterized by their inconvenience costs $\theta_k, k \in \mathcal{K}_{i,j}$, and a network-wide population of on-demand EVs characterized by their inconvenience costs $\theta_{\ell}, \ell \in \mathcal{L}$.
We denote the mobile storage capacity provided by commuter EVs with fixed routes by $\mathbf{S}^{\text{fix}}$, and the capacity provided by on-demand EVs with flexible routes by $\mathbf{S}^{\text{flex}}$. The total mobile storage capacity on all routes is represented by $\mathbf{S}$, and includes storage capacity from commuter and on-demand EVs. The two types of EVs provide the same service and are interchangeable in terms of value generated, but have different inconvenience, travel and battery degradation costs. In order to define the optimal storage service for each of the solution concepts, we partition the population of commuter EVs on each route into two sets: those which provide service ($\mathcal{K}^+_{i,j}$) and those which don't ($\mathcal{K}^-_{i,j}$). These sets are determined by a route specific inconvenience cost threshold $\thetabar_{i,j}$. We also partition the network wide population of on-demand EVs into those which provide service on any route ($\mathcal{L}^+$) and those which don't ($\mathcal{L}^-$). These sets are determined by a network-wide inconvenience cost threshold $\thetabar$. 

\subsection{Benchmark: social welfare maximizing operator}

\makeatletter
\newcommand{\vast}{\bBigg@{4}}
\newcommand{\Vast}{\bBigg@{5}}
\makeatother
To maximize the social welfare, a central operator solves the following problem:
\begin{subequations}\label{eq:sw-joint}
\begin{align}
    \min_{\substack{\mathbf{S}^{\text{fix}}, \mathbf{S}^{\text{flex}},\\ \boldsymbol{\thetabar}^{\text{fix}}, \thetabar^{\text{flex}} }} \:  \Vast( \:  \begin{split} & J(\mathbf{S}) + \sum_{i,j}  \int_{\theta_k \leq  \thetabar^{\text{fix}}_{i,j}} (\theta_k + \kappa) \,\mathrm{d}\, F_{i,j}(\theta_k)  \\
    & \: \: + \int_{\theta_\ell \le \thetabar^{\text{flex}}}  \theta_\ell \,\mathrm{d}\, F(\theta_\ell) + \sum_{i,j} \kappa_{i,j}S^{\text{flex}}_{i,j} \end{split} \Vast)  \\
    \text{s.t.} \quad & \mathbf{S}^{\text{flex}} \geq 0, \quad \mathbf{1}^\top \mathbf{S}^{\text{flex}} = F(\thetabar^{\text{flex}}) ,\\
    & S^{\text{fix}}_{i,j} = F_{i,j}(\thetabar^{\text{fix}}_{i,j}), \quad i,j \in \mathcal N,
\end{align}
\end{subequations}
where $\mathbf{S} = \mathbf{S}^{\text{fix}} + \mathbf{S}^{\text{flex}}$. From the perspective of the power system operator, it does not matter whether mobile storage capacity comes from commuter EVs ($\mathbf{S}^{\text{fix}}$) or on-demand EVs ($\mathbf{S}^{\text{flex}}$), and we can use the aggregate mobile storage capacity for our modeling.
From \cite{agwan2021marginal}, we know that the marginal value of adding mobile storage on a route is 
\begin{equation}
    \nabla_{S^{\text{fix}}_{i,j}} J(\mathbf{S}) = \nabla_{S^{\text{flex}}_{i,j}} J(\mathbf{S}) = - (\lambda_j^{(2)}(\mathbf{S}) - \lambda_i^{(1)}(\mathbf{S}))_+,
\end{equation}
which is the same for both commuter and on-demand EVs. The operator will add mobile storage capacity to maximize social welfare, i.e. will increase mobile storage capacity on a route as long as the marginal value is greater than or equal to the cost for either type of EV, i.e.
\begin{subequations}
\begin{align}
    \lambda_j^{(2)}(\mathbf{S}) - \lambda_i^{(1)}(\mathbf{S}) & \geq \thetabar^{\text{fix}}_{i,j} + \kappa \\
    \lambda_j^{(2)}(\mathbf{S}) - \lambda_i^{(1)}(\mathbf{S}) & \geq \thetabar^{\text{flex}} + \kappa_{i,j}.
\end{align}
\end{subequations}
The operator will prioritize dispatching on-demand EVs to the route with the greatest marginal increase in social welfare. We can ignore the positive part operator in this expression, since the sum of costs on the right hand side is non-negative by definition.
\begin{lemma}\label{lem:sw-both}
The inconvenience cost thresholds for the socially optimal storage operation of commuter and on-demand EVs are given by the joint solution of \eqref{eq:theta-sw} and \eqref{eq:thetabar-sw-flex}, where $i^*, j^*$ are defined as in \eqref{eq:flex-sw-route} and prices are determined by the aggregate mobile storage capacity: 
\begin{subequations} 
    \begin{align}
    & \mathbf{S}^{\text{sw}} = \mathbf{S}^{\text{fix, sw}} + \mathbf{S}^{\text{flex, sw}} \\
     & \mathbf{1}^\top \mathbf{S}^{\text{flex, sw}} = F(\thetabar^{\text{flex, sw}}), \\
     & \rev{ S_{i,j}^{\text{fix, sw}} = F_{i,j} (\thetabar_{i,j}^{\text{fix, sw}})}, &  i,j \in \mathcal N, \\
     & \thetabar^{\text{flex, sw}} = \lambda_{j^*}^{(2)}(\mathbf{S}^{\text{sw}}) - \lambda_{i^*}^{(1)}(\mathbf{S}^{\text{sw}})  - \kappa_{i^*,j^*}, \\
     & \thetabar_{i,j}^{\text{fix, sw}} = \lambda_j^{(2)}(\mathbf{S}^{\text{sw}}) - \lambda_i^{(1)}(\mathbf{S}^{\text{sw}}) - \kappa,  &  i, j \in \mathcal N. \label{eq:both-theta-fix-sw} 
    \end{align}
\end{subequations}
\end{lemma}
\begin{proof}
The marginal increase in social welfare upon adding commuter EVs is given by \eqref{eq:sw-inc-fixed}, and the operator will add mobile storage using commuter EVs along route $i \rightarrow j$ as long as the marginal value is positive.
Similarly, the socially optimal decision for on-demand EVs is to add mobile storage as long as the marginal value is positive as discussed in the proof for Lemma \ref{lem:sw-flex}. Commuter EVs and on-demand EVs interact with each other through their effect on the electricity prices $\lambda$. The marginal increases in social welfare on adding mobile storage through commuter EVs \eqref{eq:sw-inc-fixed} or on-demand EVs \eqref{eq:inc-sw-flex} are only related through prices which depend on aggregate mobile storage capacity, and jointly solving the set of equations in Lemma \ref{lem:sw-both} will resolve the interdependencies and give us the socially optimal solution.
\end{proof}

\subsection{Joint Nash equilibrium}
Consider the situation where all of the EVs are operated independently irrespective of their type. Each EV driver participates in an EV battery sharing game, and makes an independent decision to provide mobile storage service and chose a route  (for on-demand EVs) in order to maximize $\pi_{k}(s_k, \mathbf{S})$ or $\pi_\ell(\mathbf{s}_\ell, \mathbf{S})$ as appropriate. At the equilibrium, there will be a combination of commuter and on-demand mobile storage on each route. No commuter EV should be better off if it switches from $\mathcal{K}^+_{i,j}$ to $\mathcal{K}^-_{i,j}$ or vice versa (characterized by $\thetabar_{i,j}^{\text{fix, NE}}$). Similarly, no on-demand EV should be better off switching from $\mathcal{L}^+$ to $\mathcal{L}^-$ or vice versa (characterized by $\thetabar^{\text{flex, NE}}$), or by switching routes. The storage capacities at equilibrium are given by
\begin{align}
        \mathbf{1}^\top \mathbf{S^{\text{flex, NE}}} & =  F^{\text{flex}}(\thetabar^{\text{flex, NE}}),\\
         S^{\text{fix}}_{i,j} &  = F^{\text{fix}}_{i,j}(\thetabar_{i,j}^{\text{fix, NE}}) , & \: i, j \in \mathcal{N},
\end{align}
    where $F^{\text{flex}}(\cdot), F^{\text{fix}}_{i,j}(\cdot)$ are the cumulative distributions of inconvenience costs of on-demand and commuter EVs on that route respectively. The equilibrium decision by a commuter EV is given by 
    \begin{equation}\label{eq:s_k-fix-hybrid-NE}
        s_{k;i,j}^{\text{fix, NE}} = \begin{cases}
        1, \quad \text{if} \: \lambda_j^{(2)}(\mathbf{S^{\text{NE}}}) - \lambda_i^{(1)}(\mathbf{S^{\text{NE}}}) - \theta_k - \kappa \geq 0\\
        0, \quad \text{otherwise.}
        \end{cases}
    \end{equation}
    The equilibrium decision by an on-demand EV is given by the route choice 
    \begin{equation}\label{eq:route-hybrid-NE}
    (i^*, j^*) = \argmax_{i,j}  \lambda_j^{(2)}(\mathbf{S}^{\text{NE}}) - \lambda_i^{(1)}(\mathbf{S}^{\text{NE}}) - \kappa_{i,j} ,
\end{equation}
 and the mobile storage service provision by
 \thickmuskip=2mu
 \medmuskip=2mu
 \thinmuskip=0mu
 \begin{subequations} \label{eq:s_ell-flex-hybrid-NE}
\begin{align}
& s_{\ell;\, i,j}^{\text{flex, NE}} = 0, \quad \text{if} \: (i,j) \neq (i^*, j^*), \\
   & s_{\ell;i^*,j^*}^{\text{flex, NE}} = \begin{cases}
    1, \: \text{if} \:  \lambda_{j^*}^{(2)}(\mathbf{S}^{\text{NE}}) - \lambda_{i^*}^{(1)}(\mathbf{S}^{\text{NE}}) - \theta_\ell - \kappa_{i^*,j^*} \geq 0, \\
    0, \: \text{otherwise}.
    \end{cases} 
\end{align}
\end{subequations}
\thickmuskip=5mu
 \medmuskip=4mu
 \thinmuskip=3mu
We can characterize the Nash equilibrium by considering the exhaustive list of cases:

\begin{enumerate}
    \item For a route $i \rightarrow j$, consider $\mathcal{K}^-_{i,j}$ and $\mathcal{L}^-$ are not empty, i.e., there are some commuter and on-demand EVs not providing mobile storage service. Then the marginal payoff for either type of EV is non-positive. 
    \begin{enumerate}
        \item $S^{\text{fix}}_{i,j} \neq 0$ and $ S^{\text{flex}}_{i,j} \neq 0$;
        then the marginal payoff for both type of EVs on that route should be zero, i.e.
        \begin{subequations}
    \begin{align}
        \thetabar^{\text{flex, NE}} & = \lambda_j^{(2)}(\mathbf{S^{\text{NE}}}) - \lambda_i^{(1)}(\mathbf{S^{\text{NE}}}) - \kappa_{i,j} , \\
        \thetabar_{i,j}^{\text{fix, NE}}  & =  \lambda_j^{(2)}(\mathbf{S^{\text{NE}}}) - \lambda_i^{(1)}(\mathbf{S^{\text{NE}}})  - \kappa.
    \end{align}
    \end{subequations}

    \item $S^{\text{fix}}_{i,j} = 0$, which means
    \[
    \lambda_j^{(2)}(\mathbf{S^{\text{NE}}}) - \lambda_i^{(1)}(\mathbf{S^{\text{NE}}})  - \kappa < \textstyle\min_k \theta_{k;i,j}.
    \]
    
    \item $S^{\text{flex}} = 0$, which means 
    \[
    \lambda_j^{(2)}(\mathbf{S^{\text{NE}}}) - \lambda_i^{(1)}(\mathbf{S^{\text{NE}}})  - \kappa_{i,j} < \thetabar^{\text{flex, NE}}.
    \]
    \end{enumerate}
    
\item For a route $i \rightarrow j$, consider $\mathcal{K}^-_{i,j}$ is empty, i.e., all commuter EVs provide mobile storage service. Then $S^{\text{fix, NE}}_{i,j} = 1 $, and
\begin{equation}
    \lambda_j^{(2)}(\mathbf{S^{\text{NE}}}) - \lambda_i^{(1)}(\mathbf{S^{\text{NE}}}) - \thetabar_{i,j}^{\text{fix, NE}} - \kappa \geq 0.
\end{equation}  

\item Consider $\mathcal{L}^-$ is empty, i.e., all on-demand EVs are providing mobile storage service. Then $\sum_{i,j} S^{\text{flex, NE}}_{i,j} = 1$, and
    \begin{equation}
    \lambda_j^{(2)}(\mathbf{S^{\text{NE}}}) - \lambda_i^{(1)}(\mathbf{S^{\text{NE}}}) - \thetabar^{\text{flex, NE}} - \kappa_{i,j} \geq 0
\end{equation}
for at least one route $i \rightarrow j$ in the network.

\end{enumerate}
Except for the first situation, we cannot explicitly relate the equilibrium service by commuter and on-demand EVs.
\begin{lemma}
The inconvenience cost thresholds for the equilibrium operation of commuter and on-demand EVs are given by the joint solution of \eqref{eq:theta-NE} and \eqref{eq:thetabar-ne-flex}, where $i^*, j^*$ are defined as in \eqref{eq:flex-route-ne} and prices are determined by the aggregate mobile storage capacity $\mathbf{S}^{\text{NE}} =  \mathbf{S}^{\text{fix, NE}} + \mathbf{S}^{\text{flex, NE}}$.
\end{lemma}
\begin{proof}
 Commuter EVs and on-demand EVs interact with each other through their effect on the electricity prices $\lambda$. The payoff for either type of EV depends only on the aggregate mobile storage capacity, and jointly solving the set of equations in \eqref{eq:theta-NE} and \eqref{eq:thetabar-ne-flex} will resolve the interdependencies and give us the equilibrium operation.
\end{proof}
\begin{theorem}
Any inconvenience cost thresholds for commuter and on-demand EVs corresponding to a joint Nash equilibrium also support the social welfare.
\end{theorem}
\begin{proof}
There is a one-to-one correspondence between the equations which determine inconvenience cost thresholds for the socially optimal solution and the Nash equilibrium.  
\end{proof}



    
    

\section{Conclusion} \label{sec:conclusion}
We formulate and analyze a network EV battery sharing game, where distributed EVs provide mobile energy storage service to the grid. We model two different EV behaviors: commuter EVs which travel on fixed routes, and on-demand EVs which can travel on any route in the power network. 
Our results suggest that the NE will support social welfare in settings where each EV driver is an infinitesimally small entity (and has no market power), and when only two time periods are considered. These results are robust across any combination of the different EV types. In future work, we plan to study the impact of market power for a collection of EVs, when they are coordinated by an EV aggregator or a transportation network company. We are also interested in extending our work to settings with multiple time periods. Since it is known that storage devices across the network may complement instead of substitute each other \cite{qin2018submodularity}, our positive results for competitive EV-based mobile storage may only hold under certain network congestion patterns when more than two time slots are considered.

\bibliographystyle{ieeetr}
\bibliography{main}

\begin{thebibliography}{10}

\bibitem{agwan2021marginal}
Utkarsha Agwan, Junjie Qin, Kameshwar Poolla, and Pravin Varaiya.
\newblock Marginal value of mobile energy storage in power network.
\newblock In {\em 2021 60th IEEE Conference on Decision and Control (CDC)},
  pages 4936--4943. IEEE, 2021.

\bibitem{carli2019distributed}
Raffaele Carli, Mariagrazia Dotoli, and Vittorio Palmisano.
\newblock A distributed control approach based on game theory for the optimal
  energy scheduling of a residential microgrid with shared generation and
  storage.
\newblock In {\em 2019 IEEE 15th International Conference on Automation Science
  and Engineering (CASE)}, pages 960--965. IEEE, 2019.

\bibitem{he2021utility}
Guannan He, Jeremy Michalek, Soummya Kar, Qixin Chen, Da~Zhang, and Jay~F
  Whitacre.
\newblock Utility-scale portable energy storage systems.
\newblock {\em Joule}, 5(2):379--392, 2021.

\bibitem{IEA-EVoutlook}
{International Energy Agency}.
\newblock {Global EV Outlook 2021}.
\newblock Technical report, {IEA Paris}, Note =
  {\url{https://www.iea.org/reports/global-ev-outlook-2021}}, year = {2021}.

\bibitem{kalathil2017sharing}
Dileep Kalathil, Chenye Wu, Kameshwar Poolla, and Pravin Varaiya.
\newblock The sharing economy for the electricity storage.
\newblock {\em IEEE Transactions on Smart Grid}, 2017.

\bibitem{qin2019distributed}
Junjie Qin, Sen Li, Kameshwar Poolla, and Pravin Varaiya.
\newblock Distributed storage investment in power networks.
\newblock In {\em 2019 American Control Conference (ACC)}, pages 1579--1586.
  IEEE, 2019.

\bibitem{qin2021mobile}
Junjie Qin, Kameshwar Poolla, and Pravin Varaiya.
\newblock Mobile storage for demand charge reduction.
\newblock {\em IEEE Transactions on Intelligent Transportation Systems}, 2021.

\bibitem{qin2018submodularity}
Junjie Qin, Insoon Yang, and Ram Rajagopal.
\newblock Submodularity of storage placement optimization in power networks.
\newblock {\em IEEE Transactions on Automatic Control}, 64(8):3268--3283, 2018.

\bibitem{EPA-emissions}
{United States Environmental Protection Agency}.
\newblock {Inventory of U.S. Greenhouse Gas Emissions and Sinks}.
\newblock Technical report, 2020.
\newblock
  \url{https://www.epa.gov/ghgemissions/inventory-us-greenhouse-gas-emissions-and-sinks}.

\bibitem{GTMstorage20}
{Wood Mackenzie}.
\newblock { US energy storage market shatters records in Q3 2020}, 2020.
\newblock
  \url{https://www.woodmac.com/press-releases/us-energy-storage-market-shatters-records-in-q3-2020/}.

\bibitem{yang2020selling}
Yu~Yang, Utkarsha Agwan, Guoqiang Hu, and Costas~J Spanos.
\newblock Selling renewable utilization service to consumers via cloud energy
  storage.
\newblock {\em arXiv preprint arXiv:2012.14650}, 2020.

\bibitem{zhao2019virtual}
Dongwei Zhao, Hao Wang, Jianwei Huang, and Xiaojun Lin.
\newblock Virtual energy storage sharing and capacity allocation.
\newblock {\em IEEE transactions on smart grid}, 11(2):1112--1123, 2019.

\end{thebibliography}
\end{document}